\documentclass[aps,prl,twocolumn,showpacs, letterpaper]{revtex4}
\usepackage{amsmath, amsthm, amssymb}

\begin{document}
\title{Measuring thermodynamic length}
\author{Gavin E. Crooks}
\email{GECrooks@lbl.gov}
\affiliation{Physical Bioscience Division, Lawrence Berkeley National Laboratory, Berkeley, California 94720, USA}
\date{\today}

\begin{abstract}
Thermodynamic length is a metric distance between equilibrium thermodynamic states. Among other interesting properties, this metric asymptotically bounds the dissipation induced by a finite time transformation of a thermodynamic system. It is also connected to the Jensen-Shannon divergence, Fisher information and Rao's entropy differential metric. Therefore, thermodynamic length is of central interest in understanding matter out-of-equilibrium. In this paper, we will consider how to define thermodynamic length for a small system described by equilibrium statistical mechanics and how to measure thermodynamic length within a computer simulation. Surprisingly, Bennett's classic acceptance ratio method for measuring free energy differences also measures thermodynamic length.
\end{abstract}
\preprint{LBNL-62740}
\maketitle

\newcommand{\E}{{\mathrm E}}
\newcommand{\JS}{\mathrm{JS}}
\newcommand{\kB}{k_\mathrm{B}}

\section{Introduction}

Thermodynamic length is a natural measure of the distance between equilibrium thermodynamic states~\cite{Weinhold1975a, Ruppeiner1979, Salamon1983a, Salamon1984, Schlogl1985, Salamon1985, Nulton1985a, Nulton1985b,Janyszek1989, Mrugala1990,Brody1995},  which equips the surface of thermodynamic states with a Riemannian metric and defines the length 
 of a quasi-static transformation as the number of natural fluctuations along that path. Unlike the entropy or free energy change, which are state functions, the thermodynamic length explicitly depends on the path taken through thermodynamic state space.  Thermodynamic length is  of fundamental interest to the generalization of thermodynamics to finite time (rather than infinity slow) transformations.  Minimum distance paths are geodesics on the Riemannian manifold and minimize the dissipation for slow, but finite time transformations~\cite{Salamon1983a, Nulton1985b}. 
 These insights have been employed to optimize  fractional distillation and other thermodynamic processes~\cite{Salamon1998,Schaller2001,Nulton2002}. 

The study of thermodynamic length has largely been restricted to the field of macroscopic, endoreversible thermodynamics. However, there are deep connections between thermodynamic length, information theory and the statistical physics of small systems far-from-equilibium. In this paper we will consider the most appropriate definition of thermodynamic length for small systems and how to measure this distance in a computer simulation. These considerations reveal a surprising connection between thermodynamic length, Jensen-Shannon divergence and Bennett's acceptance ratio method for free energy calculations~\cite{Bennett1976}.  Bennett's method is an optimal measure of free energy differences, but it also indirectly places a lower bound on the thermodynamic length between neighboring thermodynamic states.

\section{Thermodynamic Length}
Consider a physical  system, possible microscopically small,  in equilibrium with a large thermal reservoir.  The configurational probability distribution is given by the Gibbs ensemble,~\cite{Callen1985}
\begin{equation}
p(x|\lambda) =   \frac{1}{Z}   e^{ -\beta H(x,\lambda)} = \frac{1}{Z} e^{- \lambda^i(t) X_i(x)}
\label{gibbs}
\end{equation}
where $x$ is the configuration, $t$ is time, $\beta=1/\kB T$ is the reciprocal temperature ($T$) of the environment in natural units, ($\kB$ is the Boltzmann constant),  $Z$ is the partition function, and  $H$  is the Hamiltonian of the system.
This total Hamiltonian is split into a collection of collective variables $X_i$ and conjugate generalized forces $\lambda^i$, 
$
\beta H = \lambda^i(t) X_i(x)
\label{betaH}
$. 
We use the Einstein convention that repeated upper/lower indices are implicitly summed. 
The sub-Hamiltonians $X$ are time-independent functions of the configurations, whereas the conjugate variables $\lambda$ are time dependent and configuration independent.  Note that
the conjugate variables  include a factor of inverse temperature.

The $\lambda$'s are the experimentally controllable parameters of the system and define the accessible thermodynamic state space. For example, in the isothermal-isobaric ensemble we have $X=\{U, V\}$ and $\lambda=\{\beta, \beta p\}$, where $U$ is the internal energy, $V$ is the volume and $p$ is the external pressure.  Modern experimental techniques have broadened the range of controllable parameters beyond those considered in standard thermodynamics.  For instance,  optical tweezers can apply a constant force to the ends of a single DNA molecule. The equilibrium description of this system includes the extension of the polymer, with the tension as conjugate variable. In computer simulations we have much greater flexibility.  The configuration functions can be rather arbitrary collective variables delineating high dimensional manifolds of equilibrium thermodynamic states.

The partition function that normalizes the probability distribution, $Z$, is directly related to the free energy $F$ (Gibbs potential), the free entropy $\psi$ (Massieu potential) and entropy $S$:
\begin{equation}
\ln Z 
\,=\, -\beta F 
\,=\, \psi 
\,=\, S -  \lambda^i \langle X_i \rangle 
\end{equation}
Angled brackets indicate an average over the appropriate equilibrium ensemble.
The first derivatives of the free entropy give the first moments of the collective variables, 
\begin{equation}
 \frac{\partial \psi}{\partial \lambda^i} 
= -\langle  X_i \rangle 
\end{equation}
and the second derivative yields the covariance matrix,
\begin{equation}
g_{ij} = \frac{\partial^2 \psi}{\partial \lambda^i \partial\lambda^j} 
=- \frac{\partial  \langle  X_i \rangle }{\partial\lambda^j} 
=  \big\langle  (X_i -\langle X_i\rangle)(X_j -\langle X_j\rangle) \big\rangle
\,.
\label{metrictensor}
\end{equation}
%
%The set of coefficients $\lambda^i$ define the relevant thermodynamic state space of the system. 
%These are the controllable parameters. 

 The covariance matrix $g_{ij}$ is positive semi-definite and varies smoothly from point to point, except at macroscopic phase transitions.  Therefore, we can use the covariance matrix as a metric tensor and naturally equip the manifold of  thermodynamic states with a Riemannian  metric.
Recall that a metric provides a measure of `distance' between points. It is a real function $d(a,b)$ such that (1)  distances are non-negative,  $d(a,b) \geq 0$ with equality if and only if $a=b$, (2) symmetric, $d(a,b) = d(b,a)$ and (3) it is generally shorter to go directly from point $a$ to $c$ than to go by way of $b$, $d(a,b) + d(b,c) \geq d(a,c)$ (The triangle inequality). Moreover, in a Riemannian metric we can measure the distance along curves connecting different points. The length of a curve parameterized by $t$, from $0$ to $\tau$, is 
\begin{equation}
\mathcal{L} = \int_0^{\tau} \sqrt{ \frac{d\lambda^i}{dt} g_{ij} \frac{d\lambda^j}{dt}  } dt
\label{TL}
\end{equation}
and the point-to-point distance is the length of the shortest curve. Curves of locally minimal  distance are called geodesics, and are the closest analogs of straight lines in a curved space.  Because of the connection to fluctuations [Eq.~(\ref{metrictensor})] the length of curves in thermodynamic state space are measured by the number of natural fluctuations along the path. The larger the fluctuations the closer points are together~\cite{Wootters1981,Andresen1988}.

Originally, Weinhold~\cite{Weinhold1975a} defined the thermodynamic length $\mathcal L$ using the second derivatives of the internal energy $U(S,V,N)$ with respect to the extensive variables as a metric tensor, and by Ruppeiner~\cite{Ruppeiner1979} using the corresponding derivatives of the entropy, $S(U,V,N)$. Using intensive variable derivatives of the free energy was first discussed by Schl\"ogl~\cite{Schlogl1985, Mrugala1990,Brody1995}. For macroscopic thermodynamic systems these different definitions of the metric are essentially equivalent~\cite{Salamon1984, Schlogl1985}, analogously to the macroscopic equivalence of ensembles.  
However, in small systems these metrics are in general different and the Weinhold and Ruppeiner metrics may not exist, since the second derivatives of the entropy  and entropy are not guaranteed to be positive. The definition adopted in this paper [Eqs.~(\ref{metrictensor})], essentially that of Schl\"ogl, does not require the thermodynamic limit.

Moreover, with this definition we can make an important connection to statistical estimation theory, since  the thermodynamic metric tensor Eq.~(\ref{metrictensor}) is then identical to the Fisher information matrix~\cite{Cover1991}.
\begin{align}
g_{ij}(\lambda) &=\sum_x p(x) \frac{\partial \ln p(x)}{\partial\lambda^i}\frac{\partial \ln p(x)}{\partial\lambda^j} 
\label{fisher}
\\&= \sum_x p(x)  (X_i + \frac{\partial \psi}{\partial\lambda^i} ) (X_j+ \frac{\partial \psi}{\partial\lambda^j} ) 
\nonumber
\\&=  \big\langle  (X_i -\langle X_i\rangle)(X_j -\langle X_j\rangle) \big\rangle 
\nonumber
\end{align}
According to the Cram\'er-Rao inequality the variance of any unbiased estimator is at least as high as the inverse of the Fisher information~\cite{Cover1991}.

In 1945 Rao introduced the `entropy differential metric', the distance between two distributions arising from the Riemannian metric over the parameter space with the Fisher information metric tensor~\cite{Rao1945,Burbea1982}. This entropy differential metric is identical to the thermodynamic length when, as here, the variables are conjugate parameters of a Gibbs ensemble~\cite{Brody1995}. Note that if we plug the Fisher information metric tensor [Eq.~(\ref{fisher})] into the curve length [Eq.~(\ref{TL})] we can rewrite the entropy differential metric as~\cite{ Wootters1981, Salamon1985}
\begin{equation}
\mathcal{L} = \int_0^{\tau} \sqrt{\sum_x \frac{1}{p(x)} \left[\frac{dp(x)}{dt}\right]^2} dt
\label{fishermetric}
\end{equation}
We should probable consider Rao's definition as more general and fundamental than the thermodynamic definition, just as the statistical definition of entropy is widely  considered more general and fundamental than the original thermodynamic definition. In particular, the entropy differential metric natural extends to the situation where the Hamiltonian is not a linear function of the control parameters, or where the system is not in thermal equilibrium.

We can also define a related quantity, the thermodynamics divergence of the path,
\begin{equation}
\mathcal{J} = \tau \int_0^{\tau} \frac{d\lambda^i}{dt} g_{ij} \frac{d\lambda^j}{dt}   dt
\end{equation}
In Riemannian geometry $\mathcal{J}/2\tau$ is called the energy, or action, of the curve, due to similarity with the kinetic energy integral in classical mechanics. The length and divergence are related by the inequality,
\begin{equation}
\mathcal{J} \geq {\mathcal L}^2
\label{length-divergence}
\end{equation}
which can be derived as a consequence of the Cauchy-Schwarz inequality
$
\int_0^{\tau} f^2 dt \int_0^{\tau} g^2 dt \geq \left[ \int_0^{\tau} f g \, dt \right]^2
$
with $g(t)=1$. The value of the divergence depends on the parametrization. The minimum value ${\mathcal L}^2$ is attained only when the integrand is a constant along the path.

Thermodynamic length and divergence  control the dissipation of finite time thermodynamic transformations as we approach  the infinity slow quasi-static limit~\cite{Salamon1983a, Salamon1985, Nulton1985b}.
Consider a  protocol that perturbs the conjugate variables of the system from $\lambda_1$ to $\lambda_N$ in a series of discrete steps~\cite{Nulton1985b}. After each step we pause and allow the system to reequilibrate.  After we get to the final thermodynamic state, we run the protocol in reverse, until we again reach the initial thermodynamic state. 

The total average change in entropy of a single step is
$
\Delta S_{\text{total}} = \Delta S_{\text{system}} + \lambda^i_{t+1} \left[  \langle X_i\rangle_{t+1} -\langle X_i\rangle_{t} \right]
$~\cite{Salamon1985,Nulton1985b}.
Thus, the hysteresis, the total average dissipation of the combined forward and backwards protocols, is 
\begin{align}
\omega&= 
\sum_{t=1}^{N-1} \left( \lambda^i_{t+1} \left[  \langle X_i\rangle_{t+1} -\langle X_i\rangle_{t} \right]
+ 
% \sum_{t=1}^{N-1}  
\lambda^i_t \left[  \langle X_i\rangle_{t} -\langle X_i\rangle_{t+1} \right]  \right)
\,,
\nonumber\\
&= \sum_{t=1}^{N-1}  \left[\lambda^i_{t+1} - \lambda^i_t \right] \left[ \langle X_i\rangle_{t+1} -\langle X_i\rangle_{t} \right]
\,,
\label{hysteresis}
\\ \nonumber
&= \sum_{t=1}^{N-1}  \Delta \lambda^i  \Delta \langle X_i \rangle 
\,,
\end{align}
which we  can also  write as
\begin{equation}
\omega
=  \frac{\tau}{N} 
\sum_{t=1}^{N-1}  \frac{\Delta \lambda^i}{\delta t}  
	\frac{\Delta \langle X_i \rangle}{\Delta \lambda^j} 
	\frac{\Delta \lambda^j}{\delta t}  \delta t
\,,
\end{equation}
where $\tau = N \delta t$. In the continuum limit  we can replace the sum by an integral and find that
\begin{equation}
\lim_{N\rightarrow\infty}   N \sum_{t=1}^{N-1}   \Delta \lambda^i  \Delta \langle X_i \rangle
= %{\mathcal J} 
   \tau 
    \int_0^{\tau}  \frac{d\lambda^i}{dt} g_{ij} \frac{d\lambda^j}{dt}   dt 
 = \mathcal J
\label{hysteresis2}
\end{equation}
 As the number of steps along a path increases we approach a reversible, quasi-static process. 
In this limit, the hysteresis scales as the thermodynamic divergence and inversely as the number of steps. 
(Note that this expression differs by a factor of 2 from Ref.~\cite{Nulton1985b} because here we have considered the hysteresis, the combined dissipation of the forward and reversed protocols, rather than the dissipation along a single direction.)
Similar reasoning relates the divergence and the hysteresis of a slow, finite time transformation~\cite{Salamon1983a}. %The same result can be obtained with more explicit probabilistic reasoning~\cite{Salamon1985}.

The asymptotic hysteresis and thermodynamic divergence of a protocol will depend on the parametrization of the path.  However, thanks to the length-divergence inequality $\mathcal{J} \geq {\mathcal L}^2$
[Eq.~(\ref{length-divergence})] we know that the minimum thermodynamic divergence  of the path is the square of the thermodynamic length.  Repeating the previous analysis, we find that the thermodynamic length is related to the cumulative root mean single-step hysteresis. 
\begin{equation}
 \lim_{N\rightarrow\infty}   N \sum_{t=1}^{N-1}  \sqrt{ \Delta \lambda^i  \Delta \langle X_i \rangle} 
= {\mathcal L} 
\label{hysteresis3}
\end{equation}
Consequently, we can locate optimal, minimal dissipation paths connecting two thermodynamic states by measuring and optimizing the thermodynamic length.

\section{Measuring Thermodynamic Length}
Thermodynamic length and divergence are clearly of fundamental interest and importance to non-equilibrium thermodynamics. Therefore, we shall consider how best to measure these quantities.
The relation between dissipation and divergence [Eqs.~(\ref{hysteresis2}) and~(\ref{hysteresis3})] suggests one obvious approach. We run equilibrium simulations at a series of  points along the path and examine the scaling of the dissipation with the number of steps.
Since length and divergence are properties of the path taken through thermodynamic state space, but  are independent of the underlying dynamics of the system, one can measure thermodynamic length in a computer simulation using whatever dynamics is most convenient, be it Metropolis Monte Carlo, Langevin dynamics or deterministically thermostated molecular dynamics. The only condition is that the chosen dynamics reproduce the correct equilibrium ensemble, Eq.~(\ref{gibbs}). 

Concretely, we must measure $\Delta \langle X_i \rangle$, the mean change of the collective variables between neighboring thermodynamic states. Given $K$ uncorrelated measurements from an equilibrated computer simulation we can estimate this value as 
\begin{align}
%\langle X_i\rangle_{t+1} -\langle X_i\rangle_{t}  
\Delta \langle X_i \rangle
&= \sum_x p(x|\lambda_{2}) X_i(x) - \sum_x p(x|\lambda_1) X_i(x)  
\label{perturbation}
\\
&= \sum_x p(x|\lambda_1)X_i(x)   \left(\frac{p(x|\lambda_{2})}{p(x|\lambda_1)} -1\right)
\nonumber\\
&\approx \sum_{k=1}^K   X_{i,1,n}  (\exp\left( \Delta\psi_{12} -  (\lambda^j_2 - \lambda^j_1)  X_{j,1,k}  \right) -1 )
\nonumber
\end{align}
In the second line we rewrite the difference of the means as the mean difference. 
(We should not estimate the difference of the mean directly  since this will lead  to large statistical errors that will become larger as the number of steps increases.) The final line follows from the definition of the Gibbs ensemble, Eq.~(\ref{gibbs}). 
Here, $X_{i,t,k}$ is $k$th measurement of the $i$th collective variable, $X_i(x)$ taken from an equilibrium system defined by the conjugate variables $\lambda_t$, and $\Delta\psi_{12} =  \psi_{2} -\psi_{1}$ is the difference in free entropy.

To employ Eq.~(\ref{perturbation}) we need to know the free entropy change, $\Delta\psi_{12}$, which can be optimally estimated using  Bennett's acceptance ratio method~\cite{Bennett1976, Shirts2003, Maragakis2006}. Given $K$ measurements from each of two neighboring states, $X_{i,1,k}$ and $X_{i,2,k}$ the log likelihood $\ell$ that the free entropy has a particular value is~\cite{Shirts2003, Maragakis2006}
\begin{align}
\ell(\Delta \psi_{12}) & =
\nonumber
 \frac{1}{K} \sum_{k=1}^K \ln \frac{1}{1+\exp\left( -\Delta \psi_{12} + (\lambda^i_2 - \lambda^i_1) X_{i,1,k}  \right)} 
 \\
+
\frac{1}{K} &\sum_{k=1}^{K} \ln \frac{1}{1+\exp\left(- \Delta \psi_{21} + (\lambda^i_1 - \lambda^i_2) X_{i,2,k}  \right) }
    \label{likelihood}
\end{align}
and the Bennett optimal estimate of $\Delta \psi_{12}$ maximizes this likelihood. (See~\cite{Shirts2003}
for a clear and concise exposition of this result.)

Rather than using  this free entropy measurement to estimate the mean change in the collective variables using Eq.~(\ref{perturbation}), we will instead show that the Bennett likelihood is directly related to the thermodynamic divergence. If we insert the Gibbs ensemble [Eq.~(\ref{gibbs})] into the log likelihood, then in the large sample limit, we find that the likelihood scales as 
\begin{equation}
\ell (\Delta \psi_{12})  \simeq 2K\left(  \JS(p^1; p^2) - \ln 2\right) 
   \label{likelihood2}
\end{equation}
where $ \JS(p^1; p^2)$ is the Jensen-Shannon divergence,  the mean of the relative entropy of each distribution to the mean distribution~\cite{Lin1991}.
\begin{equation}
\JS(p;q) = \frac{1}{2} \sum_i p_i \ln \frac{p_i}{\frac{1}{2}(p_i+q_i)} +  \frac{1}{2}\sum_i q_i \ln \frac{q_i}{\frac{1}{2}(p_i+q_i)} 
\end{equation}
The minimum divergence is zero for identical distributions and the maximum is $\ln2$. 
The square root of the Jensen-Shannon divergence is a metric between probability distributions~\cite{Endres2003}. However, unlike a Riemannian metric, the Jensen-Shannon metric space is not an intrinsic  length space.  There may not be a mid point $b$ between points $a$ and $c$  such that $d(a,b) +d(b,c) = d(a,c)$ and consequentially we cannot naturally measure path lengths. However, on any metric space we can define a new intrinsic metric by measuring the distance along continuous paths. The Jensen-Shannon divergence between infinitesimally different distributions is~\cite{Majtey2005} 
\begin{equation}
\JS(p; p+dp) =\frac{1}{8} \sum_i \frac{(dp_i)^2}{p_i} \,.
\end{equation}
If we compare with Eq.~(\ref{fishermetric}), we can see that in the continuum limit
\begin{equation}
\mathcal{L} = \sqrt{8} \int d\sqrt{\JS} \qquad \text{and} \qquad \mathcal{J} = 8 \int d\JS \,.
\label{connection}
\end{equation}
The induced Jensen-Shannon metric is proportional to the thermodynamic (entropy differential) metric, and the induced Jensen-Shannon divergence is proportional to the thermodynamic divergence. 
Consequentially, the square root of Jensen-Shannon divergence between two thermodynamic states gives a lower bound on the thermodynamic length of any path between those same states, and the  Jensen-Shannon divergence is a lower bound to the thermodynamic divergence.

To summarize, we can measure the thermodynamic length and minimum thermodynamic divergence along a path in thermodynamics state space by adapting Bennett's method. We perform a series of equilibrium simulations along the path and find the maximum likelihood free entropy change [Eq.~(\ref{likelihood})] and Jensen-Shannon divergence [via Eq.~(\ref{likelihood2})] between neighboring ensembles. The cumulative Jensen-Shannon  metric along the path provides a lower bound to the thermodynamic length [Eq.~(\ref{connection})] and  a lower bound to the minimum divergence of the path [via Eq.~(\ref{length-divergence})]. This procedure is then repeated with finer discretizations of the path, until the estimates of divergence and length converge.

\begin{acknowledgments} 
This research was supported by the Department of Energy, under contract
DE-AC02-05CH11231.
\end{acknowledgments}

\bibliography{GECLibrary}

\begin{thebibliography}{26}
\expandafter\ifx\csname natexlab\endcsname\relax\def\natexlab#1{#1}\fi
\expandafter\ifx\csname bibnamefont\endcsname\relax
  \def\bibnamefont#1{#1}\fi
\expandafter\ifx\csname bibfnamefont\endcsname\relax
  \def\bibfnamefont#1{#1}\fi
\expandafter\ifx\csname citenamefont\endcsname\relax
  \def\citenamefont#1{#1}\fi
\expandafter\ifx\csname url\endcsname\relax
  \def\url#1{\texttt{#1}}\fi
\expandafter\ifx\csname urlprefix\endcsname\relax\def\urlprefix{URL }\fi
\providecommand{\bibinfo}[2]{#2}
\providecommand{\eprint}[2][]{\url{#2}}

\bibitem[{\citenamefont{Weinhold}(1975)}]{Weinhold1975a}
\bibinfo{author}{\bibfnamefont{F.}~\bibnamefont{Weinhold}},
  \bibinfo{journal}{J. Chem. Phys.} \textbf{\bibinfo{volume}{63}},
  \bibinfo{pages}{2479} (\bibinfo{year}{1975}).

\bibitem[{\citenamefont{Ruppeiner}(1979)}]{Ruppeiner1979}
\bibinfo{author}{\bibfnamefont{G.}~\bibnamefont{Ruppeiner}},
  \bibinfo{journal}{Phys. Rev. A} \textbf{\bibinfo{volume}{20}},
  \bibinfo{pages}{1608} (\bibinfo{year}{1979}).

\bibitem[{\citenamefont{Salamon and Berry}(1983)}]{Salamon1983a}
\bibinfo{author}{\bibfnamefont{P.}~\bibnamefont{Salamon}} \bibnamefont{and}
  \bibinfo{author}{\bibfnamefont{R.~S.} \bibnamefont{Berry}},
  \bibinfo{journal}{Phys. Rev. Lett.} \textbf{\bibinfo{volume}{51}},
  \bibinfo{pages}{1127} (\bibinfo{year}{1983}).

\bibitem[{\citenamefont{Salamon et~al.}(1984)\citenamefont{Salamon, Nulton, and
  Ihrig}}]{Salamon1984}
\bibinfo{author}{\bibfnamefont{P.}~\bibnamefont{Salamon}},
  \bibinfo{author}{\bibfnamefont{J.}~\bibnamefont{Nulton}}, \bibnamefont{and}
  \bibinfo{author}{\bibfnamefont{E.}~\bibnamefont{Ihrig}}, \bibinfo{journal}{J.
  Chem. Phys.} \textbf{\bibinfo{volume}{80}}, \bibinfo{pages}{436}
  (\bibinfo{year}{1984}).

\bibitem[{\citenamefont{Schl\"ogl}(1985)}]{Schlogl1985}
\bibinfo{author}{\bibfnamefont{F.}~\bibnamefont{Schl\"ogl}},
  \bibinfo{journal}{Z. Phys. B} \textbf{\bibinfo{volume}{59}},
  \bibinfo{pages}{449} (\bibinfo{year}{1985}).

\bibitem[{\citenamefont{Salamon et~al.}(1985)\citenamefont{Salamon, Nulton, and
  Berry}}]{Salamon1985}
\bibinfo{author}{\bibfnamefont{P.}~\bibnamefont{Salamon}},
  \bibinfo{author}{\bibfnamefont{J.~D.} \bibnamefont{Nulton}},
  \bibnamefont{and} \bibinfo{author}{\bibfnamefont{R.~S.} \bibnamefont{Berry}},
  \bibinfo{journal}{J. Chem. Phys.} \textbf{\bibinfo{volume}{82}},
  \bibinfo{pages}{2433} (\bibinfo{year}{1985}).

\bibitem[{\citenamefont{Nulton and Salamon}(1985)}]{Nulton1985a}
\bibinfo{author}{\bibfnamefont{J.~D.} \bibnamefont{Nulton}} \bibnamefont{and}
  \bibinfo{author}{\bibfnamefont{P.}~\bibnamefont{Salamon}},
  \bibinfo{journal}{Phys. Rev. A} \textbf{\bibinfo{volume}{31}},
  \bibinfo{pages}{2520} (\bibinfo{year}{1985}).

\bibitem[{\citenamefont{Nulton et~al.}(1985)\citenamefont{Nulton, Salamon,
  Andresen, and Anmin}}]{Nulton1985b}
\bibinfo{author}{\bibfnamefont{J.}~\bibnamefont{Nulton}},
  \bibinfo{author}{\bibfnamefont{P.}~\bibnamefont{Salamon}},
  \bibinfo{author}{\bibfnamefont{B.}~\bibnamefont{Andresen}}, \bibnamefont{and}
  \bibinfo{author}{\bibfnamefont{Q.}~\bibnamefont{Anmin}}, \bibinfo{journal}{J.
  Chem. Phys.} \textbf{\bibinfo{volume}{83}}, \bibinfo{pages}{334}
  (\bibinfo{year}{1985}).

\bibitem[{\citenamefont{Janyszek and Mruga\l{}a}(1989)}]{Janyszek1989}
\bibinfo{author}{\bibfnamefont{H.}~\bibnamefont{Janyszek}} \bibnamefont{and}
  \bibinfo{author}{\bibfnamefont{R.}~\bibnamefont{Mruga\l{}a}},
  \bibinfo{journal}{Phys. Rev. A} \textbf{\bibinfo{volume}{39}},
  \bibinfo{pages}{6515} (\bibinfo{year}{1989}).

\bibitem[{\citenamefont{Mruga\l{}a et~al.}(1990)\citenamefont{Mruga\l{}a,
  Nulton, Sch\"on, and Salamon}}]{Mrugala1990}
\bibinfo{author}{\bibfnamefont{R.}~\bibnamefont{Mruga\l{}a}},
  \bibinfo{author}{\bibfnamefont{J.~D.} \bibnamefont{Nulton}},
  \bibinfo{author}{\bibfnamefont{J.~C.} \bibnamefont{Sch\"on}},
  \bibnamefont{and} \bibinfo{author}{\bibfnamefont{P.}~\bibnamefont{Salamon}},
  \bibinfo{journal}{Phys. Rev. A} \textbf{\bibinfo{volume}{41}},
  \bibinfo{pages}{3156} (\bibinfo{year}{1990}).

\bibitem[{\citenamefont{Brody and Rivier}(1995)}]{Brody1995}
\bibinfo{author}{\bibfnamefont{D.}~\bibnamefont{Brody}} \bibnamefont{and}
  \bibinfo{author}{\bibfnamefont{N.}~\bibnamefont{Rivier}},
  \bibinfo{journal}{Phys. Rev. E} \textbf{\bibinfo{volume}{51}},
  \bibinfo{pages}{1006} (\bibinfo{year}{1995}).

\bibitem[{\citenamefont{Salamon and Nulton}(1998)}]{Salamon1998}
\bibinfo{author}{\bibfnamefont{P.}~\bibnamefont{Salamon}} \bibnamefont{and}
  \bibinfo{author}{\bibfnamefont{J.~D.} \bibnamefont{Nulton}},
  \bibinfo{journal}{Europhys. Lett.} \textbf{\bibinfo{volume}{42}},
  \bibinfo{pages}{571} (\bibinfo{year}{1998}).

\bibitem[{\citenamefont{Schaller et~al.}(2001)\citenamefont{Schaller, Hoffmann,
  Siragusa, Salamon, and Andresen}}]{Schaller2001}
\bibinfo{author}{\bibfnamefont{M.}~\bibnamefont{Schaller}},
  \bibinfo{author}{\bibfnamefont{K.~H.} \bibnamefont{Hoffmann}},
  \bibinfo{author}{\bibfnamefont{G.}~\bibnamefont{Siragusa}},
  \bibinfo{author}{\bibfnamefont{P.}~\bibnamefont{Salamon}}, \bibnamefont{and}
  \bibinfo{author}{\bibfnamefont{B.}~\bibnamefont{Andresen}},
  \bibinfo{journal}{Comp. Chem. Eng.} \textbf{\bibinfo{volume}{25}},
  \bibinfo{pages}{1537} (\bibinfo{year}{2001}).

\bibitem[{\citenamefont{Nulton and Salamon}(2002)}]{Nulton2002}
\bibinfo{author}{\bibfnamefont{J.~D.} \bibnamefont{Nulton}} \bibnamefont{and}
  \bibinfo{author}{\bibfnamefont{P.}~\bibnamefont{Salamon}},
  \bibinfo{journal}{J. Non-Equilib. Thermodyn.} \textbf{\bibinfo{volume}{27}},
  \bibinfo{pages}{271} (\bibinfo{year}{2002}).

\bibitem[{\citenamefont{Bennett}(1976)}]{Bennett1976}
\bibinfo{author}{\bibfnamefont{C.~H.} \bibnamefont{Bennett}},
  \bibinfo{journal}{J. Comput. Phys.} \textbf{\bibinfo{volume}{22}},
  \bibinfo{pages}{245} (\bibinfo{year}{1976}).

\bibitem[{\citenamefont{Callen}(1985)}]{Callen1985}
\bibinfo{author}{\bibfnamefont{H.~B.} \bibnamefont{Callen}},
  \emph{\bibinfo{title}{Thermodynamics and an Introduction to
  Thermostatistics}} (\bibinfo{publisher}{Wiley}, \bibinfo{address}{New York},
  \bibinfo{year}{1985}), \bibinfo{edition}{2nd} ed.

\bibitem[{\citenamefont{Wootters}(1981)}]{Wootters1981}
\bibinfo{author}{\bibfnamefont{W.~K.} \bibnamefont{Wootters}},
  \bibinfo{journal}{Phys. Rev. D} \textbf{\bibinfo{volume}{23}},
  \bibinfo{pages}{357} (\bibinfo{year}{1981}).

\bibitem[{\citenamefont{Andresen et~al.}(1988)\citenamefont{Andresen, Berry,
  Gilmore, Ihrig, and Salamon}}]{Andresen1988}
\bibinfo{author}{\bibfnamefont{B.}~\bibnamefont{Andresen}},
  \bibinfo{author}{\bibfnamefont{R.~S.} \bibnamefont{Berry}},
  \bibinfo{author}{\bibfnamefont{R.}~\bibnamefont{Gilmore}},
  \bibinfo{author}{\bibfnamefont{E.}~\bibnamefont{Ihrig}}, \bibnamefont{and}
  \bibinfo{author}{\bibfnamefont{P.}~\bibnamefont{Salamon}},
  \bibinfo{journal}{Phys. Rev. A} \textbf{\bibinfo{volume}{37}},
  \bibinfo{pages}{845} (\bibinfo{year}{1988}).

\bibitem[{\citenamefont{Cover and Thomas}(1991)}]{Cover1991}
\bibinfo{author}{\bibfnamefont{T.~M.} \bibnamefont{Cover}} \bibnamefont{and}
  \bibinfo{author}{\bibfnamefont{J.~A.} \bibnamefont{Thomas}},
  \emph{\bibinfo{title}{Elements of Information Theory}}
  (\bibinfo{publisher}{Wiley}, \bibinfo{address}{New York},
  \bibinfo{year}{1991}).

\bibitem[{\citenamefont{Rao}(1945)}]{Rao1945}
\bibinfo{author}{\bibfnamefont{C.~R.} \bibnamefont{Rao}},
  \bibinfo{journal}{Bull. Calcutta Math. Soc.} \textbf{\bibinfo{volume}{37}},
  \bibinfo{pages}{81} (\bibinfo{year}{1945}).

\bibitem[{\citenamefont{Burbea and Rao}(1982)}]{Burbea1982}
\bibinfo{author}{\bibfnamefont{J.}~\bibnamefont{Burbea}} \bibnamefont{and}
  \bibinfo{author}{\bibfnamefont{C.~R.} \bibnamefont{Rao}},
  \bibinfo{journal}{J. Multivariate Anal.} \textbf{\bibinfo{volume}{12}},
  \bibinfo{pages}{575} (\bibinfo{year}{1982}).

\bibitem[{\citenamefont{Shirts et~al.}(2003)\citenamefont{Shirts, Bair, Hooker,
  and Pande}}]{Shirts2003}
\bibinfo{author}{\bibfnamefont{M.~R.} \bibnamefont{Shirts}},
  \bibinfo{author}{\bibfnamefont{E.}~\bibnamefont{Bair}},
  \bibinfo{author}{\bibfnamefont{G.}~\bibnamefont{Hooker}}, \bibnamefont{and}
  \bibinfo{author}{\bibfnamefont{V.~S.} \bibnamefont{Pande}},
  \bibinfo{journal}{Phys. Rev. Lett.} \textbf{\bibinfo{volume}{91}},
  \bibinfo{pages}{140601} (\bibinfo{year}{2003}).

\bibitem[{\citenamefont{Maragakis et~al.}(2006)\citenamefont{Maragakis,
  Spichty, and Karplus}}]{Maragakis2006}
\bibinfo{author}{\bibfnamefont{P.}~\bibnamefont{Maragakis}},
  \bibinfo{author}{\bibfnamefont{M.}~\bibnamefont{Spichty}}, \bibnamefont{and}
  \bibinfo{author}{\bibfnamefont{M.}~\bibnamefont{Karplus}},
  \bibinfo{journal}{Phys. Rev. Lett.} \textbf{\bibinfo{volume}{96}},
  \bibinfo{pages}{100602} (\bibinfo{year}{2006}).

\bibitem[{\citenamefont{Lin}(1991)}]{Lin1991}
\bibinfo{author}{\bibfnamefont{J.}~\bibnamefont{Lin}}, \bibinfo{journal}{IEEE
  Trans. Info. Theory} \textbf{\bibinfo{volume}{37}}, \bibinfo{pages}{145}
  (\bibinfo{year}{1991}).

\bibitem[{\citenamefont{Endres}(2003)}]{Endres2003}
\bibinfo{author}{\bibfnamefont{J.}~\bibnamefont{Endres},
  \bibfnamefont{D.M.~Schindelin}}, \bibinfo{journal}{IEEE Trans. Info. Theory}
  \textbf{\bibinfo{volume}{49}}, \bibinfo{pages}{1858} (\bibinfo{year}{2003}).

\bibitem[{\citenamefont{Majtey et~al.}(2005)\citenamefont{Majtey, Lamberti,
  Martin, and Plastino}}]{Majtey2005}
\bibinfo{author}{\bibfnamefont{A.}~\bibnamefont{Majtey}},
  \bibinfo{author}{\bibfnamefont{P.~W.} \bibnamefont{Lamberti}},
  \bibinfo{author}{\bibfnamefont{M.~T.} \bibnamefont{Martin}},
  \bibnamefont{and} \bibinfo{author}{\bibfnamefont{A.}~\bibnamefont{Plastino}},
  \bibinfo{journal}{Eur. Phys. J. D} \textbf{\bibinfo{volume}{32}},
  \bibinfo{pages}{413} (\bibinfo{year}{2005}).

\end{thebibliography}

 \end{document}